\documentclass[11pt]{article}
\pdfoutput=1  % force pdfLaTeX on arXiv (PNG figures)

\usepackage[final]{acl}

\usepackage{times}
\usepackage{latexsym}

\usepackage[T1]{fontenc}
\usepackage[utf8]{inputenc}

\usepackage{microtype}

\usepackage{inconsolata}

\usepackage{graphicx}

\usepackage{multirow}
\usepackage{tablefootnote}
\usepackage{makecell}
\usepackage{amsmath}
\usepackage{amssymb}

\title{KanAdapter: A Kolmogorov-Arnold Network-based Plug-and-Play Module for Efficient Fine-tuning of Foundation Speech Models}

\author{
  Phuong Tuan Dat\textsuperscript{1} \quad
  Phuong Khai Minh\textsuperscript{2} \quad
  Tran Huy Dat\textsuperscript{3} \\
  \textsuperscript{1}\textit{Dept. of Electrical and Computer Engineering, National University of Singapore} \\
  \textsuperscript{2}\textit{School of Computing, National University of Singapore} \\
  \textsuperscript{3}\textit{Institute of Advanced Intelligence and Computing (IAIC), A$^*$STAR} \\
  \texttt{\{phuongtuandat, pkhaiminh02\}@u.nus.edu} \quad
  \texttt{Tran\_Huy\_Dat@a-star.edu.sg} \\
}

\begin{document}
\maketitle

\begin{abstract}
Fully fine-tuning self-supervised learning (SSL) speech models for downstream tasks is computationally prohibitive, and existing parameter-efficient fine-tuning approaches predominantly rely on MLP-based adapters whose fixed activation functions limit their representational expressiveness under tight parameter budgets. We propose \textbf{KanAdapter}, a lightweight adapter framework that replaces conventional MLP bottlenecks with Group-Rational Kolmogorov-Arnold Network (GR-KAN) modules for more expressive and parameter-efficient adaptation. Following a parallel bottleneck design, KanAdapter inserts trainable GR-KAN branches alongside frozen Transformer encoder blocks and leverages weight transfer from pre-trained MLP layers for stable initialization. Across speaker verification, speech emotion recognition, and deepfake detection, KanAdapter achieves up to 97.5\% reduction in trainable parameters relative to full fine-tuning while remaining highly competitive, and consistently outperforms AdaptFormer under comparable parameter budgets. In continual learning, it yields up to 83.6\% error reduction over full fine-tuning and MLP-based adapters, which we attribute to the localized nature of GR-KAN's rational activations that mitigates catastrophic forgetting. To our knowledge, this is the first work to explore KAN-based modules for parameter-efficient fine-tuning of speech foundation models.
\end{abstract}

\section{Introduction}

Self-supervised learning (SSL) has become a foundational paradigm in speech processing, where models pre-trained on massive unlabeled data acquire rich, transferable representations. Representative models such as wav2vec 2.0 \cite{wav2vec2}, XLS-R \cite{xlsr}, HuBERT \cite{9585401}, and WavLM \cite{9814838} achieve state-of-the-art performance across downstream tasks including speaker verification (SV), speech emotion recognition (SER), and deepfake detection (DFD). However, fully fine-tuning these large-scale models is computationally prohibitive, often exceeding the resources available in practical deployment.

Parameter-efficient fine-tuning (PEFT) methods address this by injecting small trainable modules into the frozen backbone rather than updating all parameters, as in AdaptFormer \cite{chen2022adaptformer} and LoRA \cite{lora}. However, these adapters are predominantly built upon multilayer perceptron (MLP) modules, whose fixed activation functions limit their function-approximation capacity compared to more flexible, learnable alternatives \cite{mlp-compare1, mlp-compare2}, motivating the search for more expressive adapters.

Kolmogorov–Arnold Networks (KANs) \cite{kolmogorov} are a compelling such alternative. Grounded in the Kolmogorov–Arnold representation theorem, KANs replace the fixed activations of MLPs with learnable univariate mappings on network edges, yielding a more flexible and expressive approximation framework. They overcome several core MLP limitations in modeling complex, high-dimensional mappings, and further exhibit improved resistance to catastrophic forgetting \cite{kolmogorov} — both properties particularly desirable given the high-dimensional nature of speech and the continual learning scenarios in multi-task speech processing.

In this work, we propose KanAdapter, a lightweight KAN-based adapter framework for efficient fine-tuning of speech SSL foundation models. KanAdapter integrates Group-Rational KAN (GR-KAN) \cite{kat_transformer} modules as plug-in adapters within frozen SSL backbones, enabling expressive and parameter-efficient adaptation without modifying the pre-trained weights.

Our contributions are summarized as follows:
\begin{enumerate}
\item We propose KanAdapter, a simple yet effective GR-KAN-based adapter for parameter-efficient fine-tuning of speech SSL models. To our knowledge, this is the first work to explore KAN-based networks for efficient fine-tuning of speech foundation models.

\item Extensive experiments on speaker verification, speech emotion recognition, and deepfake detection show that KanAdapter outperforms existing fine-tuning approaches while using significantly fewer trainable parameters.

\item Continual learning evaluations show that KanAdapter exhibits substantially stronger resistance to catastrophic forgetting than both full fine-tuning and existing adapter-based methods.
\end{enumerate}

\section{Related Work}

Kolmogorov–Arnold Networks (KANs) have recently emerged as a compelling alternative to MLPs, grounded in the Kolmogorov–Arnold representation theorem. The seminal KAN works \cite{kolmogorov, kolmogorov2} instantiate this theorem using B-spline basis functions as learnable edge-wise activations, and further reveal superior continual learning relative to MLPs: the localized nature of spline bases means updates to one input region minimally interfere with representations elsewhere, naturally mitigating catastrophic forgetting.

\textbf{KANs in Computer Vision.} \citet{kan_cv_continual_learning} introduces the Kolmogorov–Arnold Classifier (KAC), replacing the linear classification head with an RBF-based KAN \cite{fastkan} to achieve a better stability–plasticity balance in Class Incremental Learning. From a PEFT perspective, \citet{loki_cvpr2025} proposes LoKi, an adapter using linear down/up-projections with a minimal-depth KAN as the intermediate nonlinearity, generalizing well across nine image and video benchmarks. \citet{kat_transformer} introduces Group Rational KAN (GR-KAN), which replaces spline bases with grouped rational functions to reduce memory and latency while retaining expressive power; as drop-in MLP feed-forward replacements in Vision Transformers, it achieves state-of-the-art results on ImageNet-1K.

\textbf{KANs in Speech Processing.} \citet{phuongtuandat} uses a GR-KAN module as a feature projector between a frozen SSL backbone and a downstream classifier, achieving state-of-the-art EERs on the ASVspoof2021 LA and DF benchmarks \cite{asv21} by training only the lightweight projector. \citet{xlsr-kanformer} proposes Kanformer, integrating KAN layers into the Conformer to replace feed-forward sub-modules, with competitive speech deepfake detection performance. \citet{kan_slu} conducts a systematic study on incorporating KAN layers into spoken language understanding pipelines, offering practical insights into where KAN modules contribute most effectively.

\textbf{PEFT.} Parameter-efficient fine-tuning (PEFT) adapts large pre-trained models while keeping most parameters frozen. Adapter-based methods \cite{nlp_peft, chen2022adaptformer} insert lightweight bottleneck modules into the backbone; LoRA \cite{lora} decomposes weight updates into low-rank matrices, equivalent to a linear bottleneck without nonlinear activation; and prompt tuning \cite{prompt_tuning} prepends learnable token embeddings. Their application to speech SSL models remains underexplored, and the few works transferring PEFT to speech \cite{peft_asc, peft_sv, peft_ser} predominantly rely on MLP-based adapters designed for a single target task. The use of KAN-based modules as more expressive adapters for parameter-efficient fine-tuning of large speech SSL models remains largely unexplored, a gap this work directly addresses through KanAdapter.

\section{Preliminary}

In this section, we briefly review the theoretical and architectural foundations
underlying KanAdapter, covering the Kolmogorov-Arnold representation theorem,
the KAN architecture, and its efficient variant Group-Rational KAN (GR-KAN).

\subsection{Kolmogorov-Arnold Representation Theorem}

The theoretical foundation of KANs rests on the Kolmogorov-Arnold representation theorem \cite{kan_theorem}, which establishes that any multivariate continuous function $f$ defined on a bounded domain can be exactly represented as a finite composition of univariate continuous functions. Formally, for a smooth function $f: [0,1]^n \rightarrow \mathbb{R}$, the theorem guarantees the existence of a decomposition of the form:

\begin{equation}
    f(x_1, \ldots, x_n) = \sum_{q=1}^{2n+1} \Phi_q \left( \sum_{p=1}^{n} \phi_{q,p}(x_p) \right)
\end{equation}

where $\phi_{q,p}: [0,1] \rightarrow \mathbb{R}$ are inner univariate functions, each operating on a single input variable $x_p$, and $\Phi_q: \mathbb{R} \rightarrow \mathbb{R}$ are outer univariate functions that aggregate the inner transformations. Notably, this representation is exact: a collection of $(2n+1)(n+1)$ univariate functions is sufficient to represent any $n$-variate continuous function without approximation error.

This decomposition can be expressed compactly in matrix form as $f(\mathbf{x}) = \Phi_{\text{out}} \circ \Phi_{\text{in}} \circ \mathbf{x}$, where the inner transformation $\Phi_{\text{in}}$ collects the univariate functions $\phi_{q,p}$ into a $(2n{+}1)\times n$ matrix and the outer transformation $\Phi_{\text{out}}$ collects the $\Phi_q$ into a row vector. This factorization reveals a fundamental insight: rather than learning a joint multivariate mapping as in conventional neural networks, any complex function can be systematically constructed from compositions of simpler single-variable transformations. This property forms the theoretical basis for KAN architectures, motivating the replacement of fixed nonlinear activations with learnable univariate functions to achieve more flexible and expressive function approximation.

\subsection{Kolmogorov-Arnold Networks}

Building upon the Kolmogorov-Arnold representation theorem, KANs \cite{kolmogorov}
generalize the notion of a KA layer by learning univariate activation functions
directly on network edges rather than at nodes as in conventional MLPs.
A single KAN layer with $d_{\text{in}}$-dimensional input and $d_{\text{out}}$-dimensional
output is formulated as:

\begin{equation}
    f(\mathbf{x}) = \Phi \circ \mathbf{x} =
    \left[ \sum_{i=1}^{d_{\text{in}}} \phi_{i,1}(x_i) \, \cdots \,
    \sum_{i=1}^{d_{\text{in}}} \phi_{i,d_{\text{out}}}(x_i) \right]
\end{equation}

where $\Phi = [\phi_{j,i}(\cdot)]$ is the $d_{\text{out}}\times d_{\text{in}}$ matrix of learnable univariate functions.
This formulation is a generalized form of the two-stage decomposition in Eq. (1),
where $\Phi = \Phi_{\text{in}} \circ \Phi_{\text{out}}$. A full KAN network is constructed
by stacking $L$ such layers: given an input $\mathbf{x}_0 \in \mathbb{R}^{d_0}$,
the network output is:

\begin{equation}
    \text{KAN}(\mathbf{x}_0) = \Phi_{L-1} \circ \Phi_{L-2} \circ \cdots \circ \Phi_0 \circ \mathbf{x}_0
\end{equation}

In practice, each univariate function $\phi$ is parameterized as a linear combination
of a SiLU activation and a B-spline basis:

\begin{equation}
    \phi(x) = w_b \, \text{silu}(x) + w_s \, \text{spline}(x)
\end{equation}

where $\text{silu}(x) = \frac{x}{1+e^{-x}}$ provides a smooth residual pathway,
and $\text{spline}(x) = \sum_i c_i B_i(x)$ is a learnable B-spline with trainable
coefficients $\{c_i\}$. This parameterization endows each edge with flexible,
data-driven nonlinearity, distinguishing KANs from MLPs where activations are
fixed and shared across all inputs.

\subsection{Group-Rational Kolmogorov-Arnold Network}

While vanilla KAN is theoretically appealing, its application to large-scale
models is hindered by three challenges: (1) B-spline functions are
computationally inefficient on modern GPUs due to their recursive nature;
(2) assigning a unique activation to every input-output pair causes parameter
count and cost to grow with spline order and grid resolution, incurring
substantially higher overhead than an equivalent MLP; and (3) the default
B-spline initialization violates the variance-preserving principle, causing
instability in deep networks \cite{kat_transformer}.

To overcome these limitations, \citet{kat_transformer} proposes Group-Rational KAN (GR-KAN),
which integrates three key designs. First, rational functions replace B-spline bases,
parameterized as a Safe Padé Activation Unit \cite{pade_activation}:

\begin{equation}
    \phi(x) = w \cdot F(x) = w \cdot \frac{a_0 + a_1 x + \cdots + a_m x^m}
    {1 + |b_1 x + \cdots + b_n x^n|}
\end{equation}

where $\{a_m\}$ and $\{b_n\}$ are learnable coefficients and $w$ is a trainable scalar. Rational functions are more GPU-friendly than B-splines, as polynomial evaluation maps naturally to parallel hardware operations. Evaluated with Horner's method, the rational function (m=5, n=4) requires only 21 FLOPs per evaluation, compared to 204 FLOPs for the equivalent B-spline.

Second, GR-KAN introduces grouped parameter sharing: the $d_{\text{in}}$ input channels are divided into $g$ groups, each containing $d_g = d_{\text{in}}/g$ channels, where the rational function coefficients $\{a_m\}$ are shared within each group while each edge retains its own scalar weight $w$. The group index of input channel $i$ is $\lfloor i/d_g \rfloor$. The forward pass is the product of a learnable weight matrix $\mathbf{W} \in \mathbb{R}^{d_{\text{in}} \times d_{\text{out}}}$ and a group-wise rational transformation $\mathbf{F}(\mathbf{x}) = [ F_{\lfloor 1/d_g \rfloor}(x_1), \cdots, F_{\lfloor d_{\text{in}}/d_g \rfloor}(x_{d_{\text{in}}}) ]$, which can equivalently be implemented as a group-wise rational transformation followed by a standard linear projection:

\begin{equation}
\begin{aligned}
    \text{GR-KAN}(\mathbf{x}) &= \mathbf{W}\mathbf{F}(\mathbf{x})^{\top} \\
    &= \text{linear}(\text{group\_rational}(\mathbf{x}))
\end{aligned}
\end{equation}

This reduces the number of unique activation functions from $d_{\text{in}} \times
d_{\text{out}}$ to $g$, bringing the total parameter count to $d_{\text{in}} \times
d_{\text{out}} + d_{\text{out}} + (m + n \times g)$ - comparable to a standard MLP
with only a constant overhead.

Third, GR-KAN adopts a variance-preserving initialization scheme by first fitting
the rational coefficients $\{a_m, b_n\}$ to approximate known activations (e.g.,
identity and Swish), then initializing $w \sim \mathcal{N}(0, \alpha / d_{\text{in}})$
where $\alpha = \mathbb{E}[F(x)^2] / \text{Var}[x]$ is estimated numerically. This
ensures stable gradient flow across layers from the outset of training.

Collectively, these properties make GR-KAN a computationally efficient,
training-stable, and expressive alternative to both vanilla KAN and standard MLP,
motivating its adoption as the core component of our proposed KanAdapter.

\section{Proposed Method}

We present the architecture of KanAdapter and describe how it is integrated
into speech SSL models for parameter-efficient fine-tuning.

\subsection{KanAdapter}

\begin{figure*}
    \centering
    \includegraphics[width=0.7\linewidth]{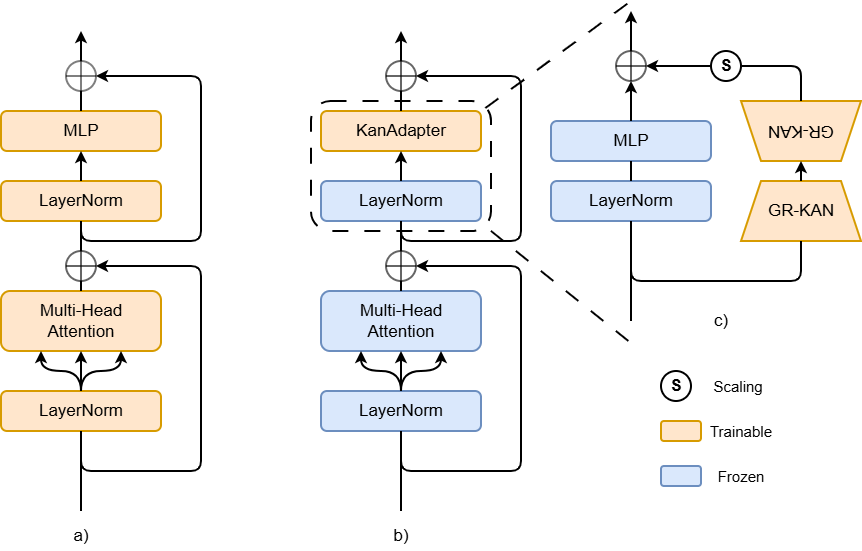}
    \caption{The proposed architecture: a) Standard Transformer layer b) AdaptKANFormer architecture c) KanAdapter.}
    \label{fig:architecture}
\end{figure*}

Inspired by AdaptFormer \cite{chen2022adaptformer}, we propose KanAdapter, a
lightweight plug-and-play adapter module designed for parameter-efficient fine-tuning
of speech SSL models. As illustrated in Figure \ref{fig:architecture}, KanAdapter
follows a parallel bottleneck design: the original frozen MLP branch of each
Transformer encoder block is preserved, while KanAdapter is inserted as an
additional trainable branch running in parallel.

\textbf{Architecture.} The overall design of KanAdapter is analogous to LoRA
\cite{lora} in structure, adopting a low-rank bottleneck with a down-projection
followed by an up-projection. However, rather than employing a standard MLP as the
intermediate nonlinear transformation, KanAdapter replaces this component with a
GR-KAN layer, leveraging its learnable rational activation functions for more
expressive feature transformation. Specifically, given an intermediate feature
$\mathbf{x}'_\ell$ from the $\ell$-th Transformer layer, KanAdapter produces the
adapted feature $\tilde{\mathbf{x}}_\ell$ as:

\begin{equation}
    \tilde{\mathbf{x}}_\ell = \text{GR-KAN}(\text{LN}(\mathbf{x}'_\ell) \cdot
    \mathbf{W}_{\text{down}}) \cdot \mathbf{W}_{\text{up}}
\end{equation}

where $\mathbf{W}_{\text{down}} \in \mathbb{R}^{d \times \hat{d}}$ and
$\mathbf{W}_{\text{up}} \in \mathbb{R}^{\hat{d} \times d}$ are the down- and
up-projection matrices, respectively, with bottleneck dimension $\hat{d} \ll d$.
The adapted feature is then fused with the frozen MLP output via a residual
connection scaled by a learnable scalar $s$:

\begin{equation}
    \mathbf{x}_\ell = \text{MLP}(\text{LN}(\mathbf{x}'_\ell)) +
    s \cdot \tilde{\mathbf{x}}_\ell + \mathbf{x}'_\ell
\end{equation}

\textbf{Fine-tuning.} During fine-tuning, the SSL backbone is kept entirely
frozen; only the KanAdapter parameters - $\mathbf{W}_{\text{down}}$,
$\mathbf{W}_{\text{up}}$, the GR-KAN rational coefficients, and the scaling
factor $s$ - are updated. This preserves the general representations learned
during pre-training while performing task-specific adaptation exclusively
through the lightweight KanAdapter branches, greatly reducing the number of
trainable parameters relative to full fine-tuning.

\section{Experiments}

We evaluate the effectiveness of KanAdapter through comprehensive experiments
across multiple speech processing tasks, including speaker verification,
speech deepfake detection, and speech emotion recognition.

\subsection{Experimental Setup}

\textbf{Pre-trained backbones:} We adopt WavLM \cite{9814838} and XLS-R \cite{xlsr} as the pre-trained backbone
models for our experiments. Specifically, for the SV and SER tasks, we employ WavLM-Large\footnote{microsoft/wavlm-large} as the encoder, while for the DFD task, we utilize XLSR-Large\footnote{facebook/wav2vec2-xls-r-300m}. We evaluate different fine-tuning strategies using three baseline architectures: WavLM-ECAPA \cite{wavlm-ecapa} for SV, WavLM with attentive statistical pooling \cite{msp_podcast} for SER, and XLSR-Conformer \cite{xlsr_conformer} for DFD.

\textbf{Initialization of KanAdapter:} The pre-trained backbone weights are loaded from upstream checkpoints and remain frozen throughout the entire fine-tuning process. For the proposed module, the down-projection layer weights are initialized from the corresponding MLP layers of the pre-trained SSL backbone, while the up-projection layer weights are initialized to zero. The rationale for zero-initializing the up-projection is to ensure that KanAdapter produces no output at the start of training, preserving the original model behavior and yielding a more stable optimization trajectory.

\textbf{Baseline methods: } We compare KanAdapter against three fine-tuning
strategies: (1) \textbf{AdaptFormer} \cite{chen2022adaptformer}: a lightweight
bottleneck adapter inserted in parallel with the frozen MLP block, employing a
down-projection, ReLU nonlinearity, and up-projection; (2) \textbf{LoRA}
\cite{lora}: a low-rank decomposition method that decouples weight updates from
the original parameters via two low-rank matrices; and (3) \textbf{Full
Fine-tuning}: all backbone parameters are optimized jointly with the downstream
task objective.

\textbf{Implementation details:} All experiments are conducted on NVIDIA A40 GPUs. For implementation, we build
upon three publicly available frameworks: WeSpeaker\footnote{\url{https://github.com/wenet-e2e/wespeaker}}
for SV, the XLSR-Conformer repository\footnote{\url{https://github.com/ErosRos/conformer-based-classifier-for-anti-spoofing.git}}
for DFD, and the MSP-Podcast codebase\footnote{\url{https://github.com/msplabresearch/MSP-Podcast_Challenge.git}}
for SER, following the default training configurations of each respective
framework. For all PEFT methods, we adopt a consistent bottleneck dimension
$\hat{d} = 128$ and apply adapter modules to every Transformer encoder layer
of the pre-trained backbone. The scaling factor is set to $s = 0.1$ for
AdaptFormer and KanAdapter, and $s = 0.5$ for LoRA.

\subsection{Datasets and Metrics}

We conduct experiments on three downstream tasks. For \textbf{Speaker Verification}, models are trained on VoxCeleb2 \cite{VoxCeleb2} (over 1M utterances) and evaluated on the VoxCeleb1 Vox1-O, Vox1-E, and Vox1-H test sets \cite{voxceleb1}, covering increasing difficulty. For \textbf{Speech Emotion Recognition}, we use MSP-Podcast \cite{msp_podcast}, collected from podcast recordings in naturalistic conditions. For \textbf{Deepfake Detection}, we train on ASVspoof2019 LA \cite{asv19} and evaluate on ASVspoof2021 LA and DF \cite{asv21} as cross-corpus test sets; for ASVspoof5 \cite{asv5}, models are trained and evaluated on its dedicated partitions. We additionally evaluate on the In-The-Wild dataset \cite{inthewild}, sourced from real-world media labeled as audio deepfakes.

Equal Error Rate (EER) \cite{eer} is employed for the SV and DFD tasks, whereas the SER task is evaluated using both F1-Macro and F1-Micro scores.

\subsection{Experimental Results}
\subsubsection{Performance Analysis on Speaker Verification}

Table \ref{tab:sv_task} presents the speaker verification results on the
VoxCeleb1 benchmark. KanAdapter achieves the most competitive performance
among all parameter-efficient methods, attaining EERs of 0.52\%, 0.50\%,
and 1.93\% on Vox1-O, Vox1-E, and Vox1-H, respectively, closely approaching
full fine-tuning (0.49\%, 0.44\%, 1.58\%) while using only 9M trainable
parameters — a 97.5\% reduction.

\begin{table}[h]
\centering
\caption{Speaker verification results on VoxCeleb1. \textbf{Bold} denotes the
best result, \underline{underline} denotes the second-best result.}
\resizebox{\columnwidth}{!}{
\begin{tabular}{lcccc}
\hline
\multirow{2}{*}{\textbf{Method}} & \textbf{\#Trainable} &
\multicolumn{3}{c}{\textbf{EER (\%)}} \\ \cline{3-5}
 & \textbf{Params} & \textbf{Vox1-O} & \textbf{Vox1-E} & \textbf{Vox1-H} \\ \hline
Full Fine-tuning & 364M & \textbf{0.49} & \textbf{0.44} & \textbf{1.58} \\
LoRA      & 4M   & 2.25          & 2.19          & 3.27           \\
AdaptFormer      & 8M   & 1.44          & 1.35          & 2.70           \\
KanAdapter       & 9M & \underline{0.52} & \underline{0.50} & \underline{1.93} \\
\hline
\end{tabular}
}
\label{tab:sv_task}
\end{table}

Although LoRA operates with the fewest parameters (4M), its linear low-rank
approximation lacks sufficient nonlinear modeling capacity, yielding
substantially higher EERs (2.25\%, 2.19\%, 3.27\%); AdaptFormer (8M) similarly
underperforms. Despite KanAdapter's modest parameter overhead, the expressive
learnable rational activations of GR-KAN enable significantly more effective
adaptation, achieving gains that far outweigh the marginal parameter increase.

\subsubsection{Performance Analysis on Speech Emotion Recognition}

Table \ref{tab:ser_task} reports speech emotion recognition results on the
MSP-Podcast dataset. KanAdapter (16M parameters) delivers the most competitive
performance among all PEFT methods, attaining F1-Macro of 0.5127 and F1-Micro
of 0.3436 on the development set — surpassing full fine-tuning on F1-Micro
despite a 95.2\% parameter reduction. On the test set, KanAdapter achieves
F1-Macro of 0.3290 and F1-Micro of 0.3180, retaining over 99\% of the full
fine-tuning performance (0.3321 and 0.3201).

\begin{table*}[h]
\centering
\caption{Speech emotion recognition results on MSP-Podcast. \textbf{Bold} denotes the best result, \underline{underline} denotes the second-best result.}
\begin{tabular}{lcccccc}
\hline
\multirow{2}{*}{\textbf{Method}} & \multirow{2}{*}{\textbf{\#Trainable Params}} &
\multicolumn{2}{c}{\textbf{Development set}} & & \multicolumn{2}{c}{\textbf{Test set}} \\
\cline{3-4} \cline{6-7}
 & & \textbf{F1-Macro} & \textbf{F1-Micro} & & \textbf{F1-Macro} & \textbf{F1-Micro} \\
\hline
Full Fine-tuning & 331M & \textbf{0.5279} & \underline{0.3433} & & \textbf{0.3321}  & \textbf{0.3201} \\
LoRA      & 8M  & 0.3126          & 0.2011             & & 0.1822 & 0.1794\\
AdaptFormer      & 14M  & 0.4123          & 0.2742             & & 0.2310 & 0.2040\\
KanAdapter       & 16M  & \underline{0.5127} & \textbf{0.3436} & & \underline{0.3290} & \underline{0.3180}\\
\hline
\end{tabular}
\label{tab:ser_task}
\end{table*}

LoRA (8M) and AdaptFormer (14M) fall substantially short on the test set
(0.1822/0.1794 and 0.2310/0.2040 F1-Macro/F1-Micro), revealing that neither
linear low-rank decomposition nor MLP-based bottlenecks provide adequate
modeling capacity for the complex affective feature distributions in SER.
The substantial gains of KanAdapter across both metrics justify its modest
parameter overhead, establishing it as the most effective parameter-efficient
adaptation strategy for this task.

\subsubsection{Performance Analysis on Deepfake Detection}

Table \ref{tab:dfd_task} presents the deepfake detection results across five
evaluation sets. Full fine-tuning establishes the performance upper bound at
the cost of 319M trainable parameters. KanAdapter reduces this to 17M — a
94.7\% reduction — while achieving the closest performance to full fine-tuning
among all PEFT methods across every evaluation set.

\begin{table}[h]
\centering
\caption{Performance of proposed method compared with baseline model on deepfake detection task. \textbf{Bold} denotes the best result, \underline{underline} denotes the second-best result.}
\resizebox{\columnwidth}{!}{
\begin{tabular}{lccccccc}
\hline
\multirow{2}{*}{\textbf{Model}} & \textbf{\#Trainable} & \multicolumn{5}{c}{\textbf{EER (\%)}}  \\ \cline{3-7}
    & \textbf{params} & \textbf{LA19} & \textbf{LA21} & \textbf{DF21} & \textbf{LA5} & \textbf{ITW}\\
\hline
Full Fine-tuning & 319M & \textbf{0.27} & \textbf{1.21} & \textbf{4.26} & \textbf{15.38} & \underline{8.34} \\
LoRA & 9M & 2.77  & 5.72 & 6.09 & 24.65 & 10.94\\
AdaptFormer & 15M & 0.53  & 4.93 & 6.51 & 19.55 & 10.21\\
KanAdapter & 17M & \underline{0.34}  & \underline{1.32}  & \underline{5.93} & \underline{18.40} & \textbf{7.96} \\
\hline
\end{tabular}
}
\label{tab:dfd_task}
\end{table}

On the in-domain LA19 set, KanAdapter attains 0.34\% EER, within 0.07\%
absolute of full fine-tuning. On the cross-corpus LA21 and DF21 sets it
achieves 1.32\% and 5.93\% EER, markedly outperforming both LoRA and
AdaptFormer, and it remains superior on LA5. The gap is particularly pronounced
on LA21, where KanAdapter achieves relative EER reductions of 76.9\% over LoRA
and 73.2\% over AdaptFormer, underscoring the advantage of expressive rational
activations over both linear decomposition and fixed-activation bottlenecks for
capturing fine-grained spoofing artifacts.

On the ITW out-of-domain benchmark, KanAdapter achieves 7.96\% EER,
surpassing not only LoRA (10.94\%) and AdaptFormer (10.21\%) but also full
fine-tuning (8.34\%) despite using a fraction of the parameters. This suggests
that the constrained adaptation of KanAdapter acts as an implicit regularizer,
yielding representations that generalize more robustly to unseen real-world
acoustic conditions.

\subsubsection{Performance Analysis on Continual Learning}

Table \ref{tab:continual_task} evaluates continual learning performance by
fine-tuning on ASVspoof5 following initial training on ASVspoof2019, measuring
the degree to which each method retains previously acquired knowledge while
adapting to new data. KanAdapter achieves the best performance across all four
evaluation sets, attaining EERs of 0.86\%, 3.59\%, 2.76\%, and 6.75\% on LA19,
LA21, DF21, and LA5, respectively.

\begin{table}[h]
\centering
\caption{Continual Learning Performance: Fine-tuning Results on ASVspoof5 Following ASVspoof19 Pre-training. \textbf{Bold} denotes the best result, \underline{underline} denotes the second-best result.}
\resizebox{\columnwidth}{!}{
\begin{tabular}{lccccc}
\hline
\multirow{2}{*}{\textbf{Model}} & \multicolumn{4}{c}{\textbf{EER (\%)}}  \\ \cline{2-5} & \textbf{LA19} & \textbf{LA21} & \textbf{DF21} & \textbf{LA5}\\ \hline
Full Fine-tuning & 5.25 & 4.74 & 6.92 & \underline{8.60} \\
LoRA & 7.19 & 9.64 & 9.28 & 20.11 \\
AdaptFormer & \underline{4.94}  & \underline{4.24} & \underline{3.38} & 11.11\\
KanAdapter & \textbf{0.86}  & \textbf{3.59}  & \textbf{2.76} & \textbf{6.75} \\
\hline
\end{tabular}
}
\label{tab:continual_task}
\end{table}

Strikingly, LoRA exhibits the worst continual learning behavior, even
degrading below full fine-tuning on all sets. This confirms that low-rank
linear decomposition is particularly susceptible to catastrophic forgetting,
as global weight updates across all layers indiscriminately overwrite previously
learned representations. Full fine-tuning and AdaptFormer similarly suffer
significant forgetting.

KanAdapter, in contrast, reduces EER by 83.6\% over full fine-tuning and
82.6\% over AdaptFormer on LA19, and consistently achieves the largest
improvements across all remaining sets. This exceptional continual learning
capability stems directly from a fundamental property of Kolmogorov-Arnold
Networks: the localized support of rational basis functions ensures that
parameter updates affect only specific input regions, naturally preventing
interference with previously acquired knowledge. This intrinsic architectural
resistance to catastrophic forgetting — absent in MLP-based methods —
represents a distinct and practically significant advantage of KanAdapter
beyond its parameter efficiency, making it particularly well-suited for
real-world deployments where models must continuously adapt to emerging
threats without forgetting prior knowledge.

\section{Ablation Study}

To further understand the behavior of KanAdapter, we conduct ablation studies examining two key aspects: the effect of bottleneck dimension on the efficiency-performance trade-off, and the generalizability of KanAdapter across different SSL backbone architectures. The ablation studies conducted in this research are all performed on the DFD task.

\subsection{Effect of Bottleneck Dimension $\hat{d}$}

\begin{figure}[h]
    \centering
    \includegraphics[width=\linewidth]{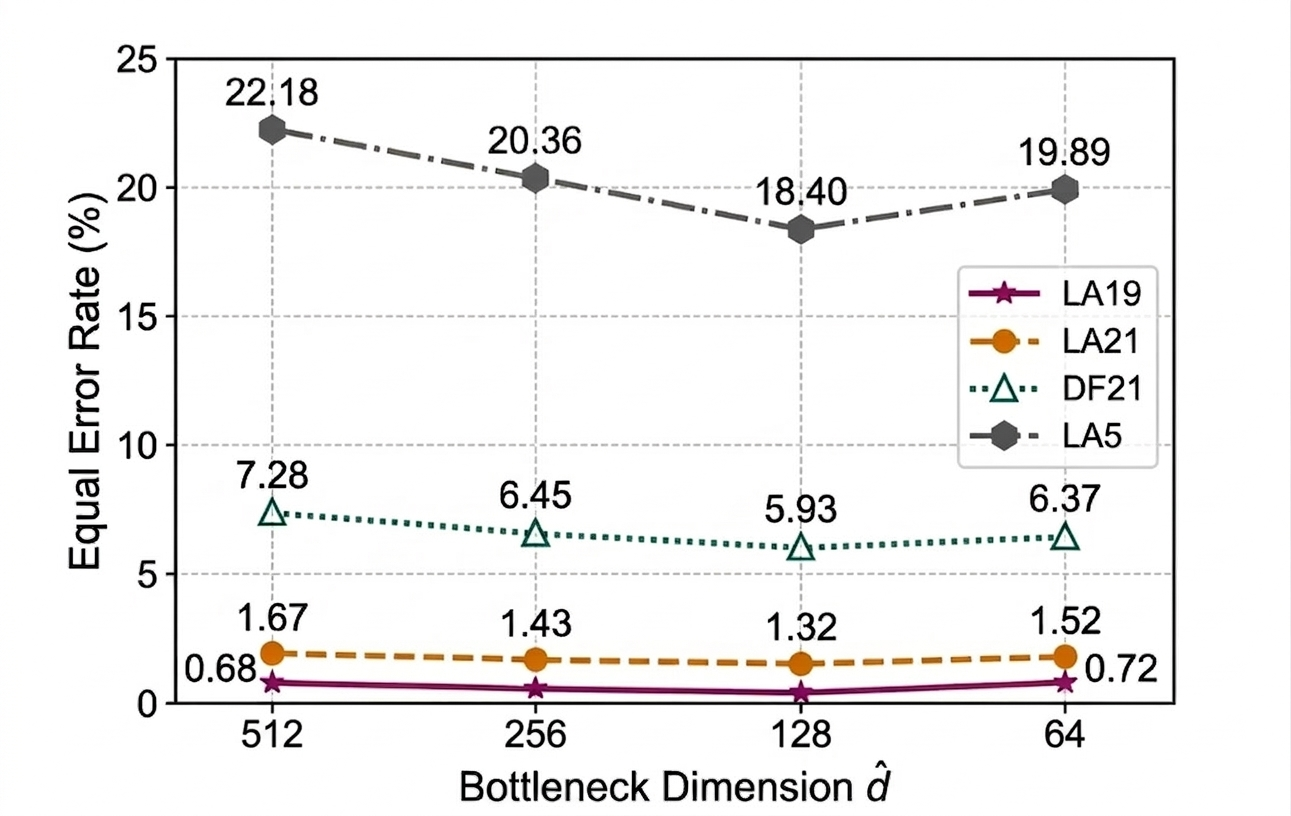}
    \caption{Performance with bottleneck dimension $\hat{d}$.}
    \label{fig:eer}
\end{figure}

The bottleneck dimension $\hat{d}$ directly controls the number of trainable
parameters introduced by KanAdapter, trading off adapter capacity against the
regularization induced by a more constrained bottleneck. We investigate this
trade-off by varying $\hat{d} \in \{512, 256, 128, 64\}$ on the deepfake
detection task and reporting EER across all four evaluation sets, as illustrated
in Figure \ref{fig:eer}.

The results indicate a consistent performance improvement as $\hat{d}$ decreases
from 512 to 128 across all evaluation sets (e.g., LA5 EER drops from 22.18\% to
18.40\%). This aligns with our earlier observation that a more constrained
bottleneck acts as an implicit regularizer, improving generalization on
cross-corpus sets. Beyond $\hat{d} = 128$, however, performance slightly degrades
at $\hat{d} = 64$, where capacity becomes the limiting factor and an excessively
small bottleneck restricts the adapter's representational power. Even at
$\hat{d} = 512$, KanAdapter remains competitive (LA19 EER of 0.68\%), showing
that the GR-KAN-based adapter retains reasonable capacity across a wide range of
configurations. We therefore set $\hat{d} = 128$ as the default, as it
consistently achieves the best performance while maintaining a favorable
parameter-efficiency trade-off.

\subsection{Generalizability Across SSL Backbones}

To assess whether KanAdapter generalizes beyond the primary XLSR-Large backbone,
we evaluate its effectiveness across three additional SSL models of varying
architectures and scales: WavLM Large (24 Transformer encoder layers, 333M
parameters), UniSpeech-SAT (12 layers, 111M parameters), and mHuBERT-147 (12
layers, 112M parameters), reporting EER on LA21 and DF21.

\begin{table}[h!]
\centering
\caption{EER (\%) result with different SSL models}
\label{tab:ssl-ablations}
\resizebox{\columnwidth}{!}{
\begin{tabular}{l|lcrr}
\hline
\textbf{SSL Model} & \textbf{Mode} &
\multicolumn{1}{c}{\textbf{\makecell{\#Trainable\\Params}}} &
\textbf{LA21} & \textbf{DF21}\\
\hline
\multirow{2}{*}{WavLM Large\tablefootnote{\url{https://huggingface.co/microsoft/wavlm-large}}}
& Full-tuning & 333M & 4.07 & 11.25\\
& KanAdapter & 17M & 4.19 & 11.76\\

\multirow{2}{*}{UniSpeech-SAT\tablefootnote{\url{https://huggingface.co/microsoft/unispeech-sat-base-plus}}}
& Full-tuning & 111M & 17.46 & 28.46\\
& KanAdapter & 10M & 19.76 & 31.45\\

\multirow{2}{*}{mHuBERT-147\tablefootnote{\url{https://huggingface.co/utter-project/mHuBERT-147}}}
& Full-tuning & 112M & 21.20 & 16.48\\
& KanAdapter & 10M & 24.56 & 20.44\\
\hline
\end{tabular}
}
\end{table}

As shown in Table \ref{tab:ssl-ablations}, KanAdapter consistently approaches
full fine-tuning performance across all three backbones while reducing trainable
parameters by 94.9\%, 91.0\%, and 91.1\%, respectively. On WavLM Large,
KanAdapter achieves EERs of 4.19\% and 11.76\% on LA21 and DF21, closely
matching full fine-tuning (4.07\% and 11.25\%) with an absolute gap of only
0.12\% and 0.51\%. This result is particularly notable given that WavLM Large
shares a similar depth (24 layers) with the primary XLSR-Large backbone,
suggesting that KanAdapter is especially effective for large-scale, deep SSL
models where the abundance of Transformer encoder layers provides sufficient
capacity for GR-KAN-based adapters to capture task-relevant representations.

Performance gaps widen for shallower models. On UniSpeech-SAT and mHuBERT-147,
both with 12 encoder layers, KanAdapter incurs larger absolute EER increases
relative to full fine-tuning, suggesting that with fewer Transformer layers
available for adapter insertion its overall adaptation capacity is more
constrained. Nevertheless, even for these smaller models KanAdapter remains a
viable alternative at a fraction of the parameter budget. These findings
highlight a practical guideline: KanAdapter yields the most favorable
efficiency-performance trade-off when applied to large, deep SSL foundation
models.

\section{Conclusion and Future Work}

In this work, we presented KanAdapter, a lightweight and plug-and-play
parameter-efficient fine-tuning framework for speech SSL foundation models,
built upon GR-KAN. By replacing the conventional MLP-based bottleneck in adapter
architectures with GR-KAN modules, KanAdapter achieves more expressive
task-specific adaptation while retaining the efficiency required for practical
deployment. Across speaker verification, speech emotion recognition, and
deepfake detection, KanAdapter delivers highly competitive performance relative
to full fine-tuning while reducing trainable parameters by up to 97.5\%, and
substantially outperforms MLP-based adapters such as AdaptFormer under
comparable parameter budgets. Continual learning evaluations further reveal
significantly stronger resistance to catastrophic forgetting (EER reductions of
up to 83.6\% on previously learned tasks), which we attribute to the localized
nature of GR-KAN's grouped rational activations. Ablation studies confirm that
KanAdapter is most effective on large, deep SSL models.

While this work focuses on speech processing, the plug-and-play nature of
KanAdapter makes it directly applicable to foundation models in other
modalities such as image and video, and to large language models, where the
continual learning properties of KAN-based architectures may offer meaningful
advantages in multi-task and lifelong learning settings. We leave these
explorations for future work.

\newpage
\section*{Limitations}

Our study has several limitations. First, the improved expressiveness of GR-KAN
over MLP-based adapters is demonstrated through downstream performance rather
than probed directly, without targeted spectral, frequency-band, or
representation-level analyses isolating \emph{why} the learnable rational
activations help. Second, its benefits are most pronounced on large, deep SSL
backbones: for shallower models such as UniSpeech-SAT and mHuBERT-147, the gap
to full fine-tuning widens. Third, our evaluation focuses on speech
classification tasks (SV, SER, and DFD); applicability to sequence-generation
tasks such as ASR, and to other modalities and large language models, remains to
be verified. Finally, continual learning is assessed only in a two-stage setting
(ASVspoof2019 followed by ASVspoof5); longer task sequences and broader
benchmarks are needed to fully characterize its resistance to catastrophic
forgetting.

% Bibliography
\bibliography{sample-base}

\end{document}